\documentclass[a4paper,fleqn,usenatbib]{mnras}

\usepackage[T1]{fontenc}
\usepackage{ae,aecompl}

\usepackage{graphicx}	
\usepackage{amsmath}	
\usepackage{amssymb}	

\title[Hot RCB Stars Secular Fading]{All Known Hot RCB Stars Are Fading Fast Over the Last Century}

\author[B. E. Schaefer]{
Bradley E. Schaefer,$^{1}$\thanks{E-mail: schaefer@lsu.edu}
\\
$^{1}$Department of Physics \& Astronomy, Louisiana State University, Baton Rouge, Louisiana, 70803, USA
}

\date{Accepted XXX. Received YYY; in original form ZZZ}

\pubyear{2015}

\begin{document}
\label{firstpage}
\pagerange{\pageref{firstpage}--\pageref{lastpage}}
\maketitle

\begin{abstract}
The R Coronae Borealis (RCB) stars are cool supergiants that display irregular and deep dips in their light curves, caused by dust formation.  There are four known hot RCB stars (DY Cen, MV Sgr, V348 Sgr, and HV 2671), with surface temperatures of 15,000--25,000 K, and prior work has suggested that three of these have secular fading in brightness.  I have tested this result by measuring century-long light curves in the Johnson B-band with modern comparison star magnitudes, and I have extended this by measuring many magnitudes over a wide time range as well as for the fourth hot RCB star.  In all four cases, the B-band magnitude of the maximum light is now fast fading.  The fading rates (in units of magnitudes per century) are 2.5 for DY Cen after 1960, 1.3 for MV Sgr, 1.3 for V348 Sgr, and 0.7 for HV 2671.  This secular fading is caused by the expected evolution of the star across the top of the HR diagram at constant luminosity, as the temperature rises and the bolometric correction changes.  For DY Cen, the brightness at maximum light is rising from 1906 to 1932, and this is caused by the temperature increase from near 5,800 to 7,500 K.  Before 1934 DY Cen had frequent dust dips, while after 1934 there are zero dust dips, so there is some apparent connection between the rising temperature and the formation of the dust.  Thus, we have watched DY Cen evolve from an ordinary RCB star up to a hot RCB star and now appearing as an extreme helium star, all in under one century.\end{abstract}




\section{Introduction}

R Coronae Borealis (RCB) stars are defined by their light curves displaying sudden and large drops in brightness with slower recoveries to the baseline level, with these events occurring randomly in time.  These spectacular dips are caused when the star forms dense dust clouds on the line of sight to Earth hiding the star.  The RCB stars are all hydrogen deficient supergiants, with various abundance anomalies, including enriched nitrogen and carbon.  RCB stars are rare, with only 76 known in our Milky Way.  The evolutionary status of RCB stars is that they might be from recent coalescences of a double white dwarf binary or from a final helium shell flash in a born-again asymptotic giant branch (AGB) star.  Clayton (1996; 2012) presents full reviews of RCB stars.

Most RCB stars are relatively cool, with surface temperatures of 5,000-7,000 K.  However, four of the known RCB stars have a greatly hotter surface temperature, from 15,000-25,000 K, and these are called the `hot RCB stars'.  The four known hot RCB stars are V348 Sgr, MV Sgr, and DY Cen in our Milky Way plus HV 2671 in the Large Magellanic Cloud.  DY Cen and MV Sgr are hydrogen deficient (just like the cool RCB stars) and mostly composed of helium, while both have relatively infrequent drops in brightness.  V348 Sgr and HV 2671 are very carbon-rich (55\% carbon, most of the rest helium), and elemental abundances like in central stars of planetary nebulae, while both have frequent episodes of brightness declines.  A plausible idea is that the two greatly different composition indicate two formation mechanisms, with DY Cen and MV Sgr simply being the `progeny' of the normal RCB stars as they heat up, and with V348 Sgr and HV 2671 being somehow formed during a final helium-shell flash on post-AGB stars.  Thus, the birth mechanism of the hot RCB stars could be either from born-again systems or white dwarf mergers.  Nevertheless, it is still unclear as to the relationship between the hot RCB stars and the cool RCB stars, as well as to other classes of stars (the born-again stars and the Wolf-Rayet central stars of planetary nebulae).  De Marco et al. (2002) present a full review of the hot RCB stars.

The key high-level science question is to understand the evolutionary state of the RCB stars, both hot and cold.  For this, a critical piece of evidence is to watch them evolve in real time.  Like for the born-again post-AGB stars (i.e., V605 Aql, Sakurai's object, and FG Sge), substantial movement across the HR diagram might be seen on time scales of a decade and a century.  For this, De Marco et al. (2002) have pointed out that the baseline levels for MV Sgr, V348 Sgr, and DY Cen are apparently fading over the last century, and this would only be from movement from right-to-left across the top of the HR diagram.

These century-long light curves have inevitable big problems for two reasons.  The light curves were compiled from magnitudes in the $V$, $m_{vis}$, and $m_{pg}$ systems, which makes for large systematic uncertainties between old and new magnitudes, with exactly this able to make apparent systematic declines when none are real.  De Marco et al. did attempt to correct for such effects.  The big effect not mentioned is that all the old photometry is always systematically in error by from 0.1 to $>$1.0 mag simply because the old standard stars had these errors.  In the days before photoelectric photometers could reach to cover faint sequences (i.e., before the late 1970s), the standard stars and comparison sequences were all calibrated by photographic transfers from the Harvard-Groningen Selected Areas and the North Polar Sequence.  These photographic transfers always had problems, due to effects like reciprocity failure, in that claimed magnitudes were always reported systematically different than what we take on the modern magnitude scales.  In general, the errors are small for stars brighter than tenth magnitude and start increasing steeply as the stars get fainter.  Thorough studies of errors for old standard fields and comparison stars are given by Sandage (2001) and Patat et al. (1997), while I have many examples of poor old calibrations (e.g., Schaefer 1994; 1995; 1996; 1998)  The old `photographic magnitude system' (i.e., $m_{pg}$) is really just a poorly calibrated B magnitude.

All the early magnitudes of hot RCB stars are photographic magnitudes, all from the Harvard plates, all from old measures, and so there are inevitably possibly-large systematic errors in their long-term light curves.  It might be that these systematic errors have created the apparent secular fading of the hot RCB stars, or it might be that the errors are such that the claim of secular fading is still correct.

In this paper, I will solve these problems for the old photometry, and answer the question of whether the hot RCB stars are fading or not.  To do this, I visited the Harvard College Observatory (HCO) in October 2015, and remeasured many magnitudes from plates dated 1896 to 1989, all on the modern Johnson B magnitude system.  For the modern portions of the light curves, I used a variety of sources from the literature and from the {\it American Association of Variable Star Observers} (AAVSO), all in the Johnson B magnitude system.  My combined light curves, all in a single uniform system, have a larger time range and many more magnitudes than given in De Marco et al.  I have further extended their work by adding the B-band light curve for the fourth and last-known hot RCB star, HV 2671.

\section{Photometry with Archival Plates}

From around 1890 to the late 1970s, a large part of astronomy was from photometry as based on photographic sky pictures.  These pictures were recorded on blue-sensitive emulsion attached to one side of a glass plate, with the plate being exposed to star light in a special holder at the focus of the telescope.  The developed emulsion had the sky being nearly transparent, while stars were round black points.  Photographic emulsions have only a small dynamic range between the sky and saturation, so star images are almost all completely saturated (i.e., black) in the centers, with only a small annulus of grey in the outer tails of the star profile.  With this, essentially, the only change in the star image as the magnitude varies is the diameter of the star image.  Given the variability from plate-to-plate and the non-linearity of the emulsion, the only way to calibrate the image-diameter-versus-magnitude relation is to use comparison stars nearby on the plate.  The procedure is to make some measure of the image diameter for the target star as well as for a sequence of nearby comparison stars with a tight spacing of magnitudes, both brighter and fainter than the target.  With this, the magnitude of the target can be measured by interpolation in the image-diameter-versus-magnitude relation as determined for each plate from the comparison sequence.

The image radii vary such that the square of the radii or the logarithm of the radii are linear with the magnitude, depending on the brightness of the star (Schaefer 1981).  In general, the full calibration curve (image radius versus magnitude) is nonlinear with either magnitude or flux.  This condition violates one of the requirements ingrained in observers with CCDs, because then magnitudes cannot be calculated from any application of the magnitude equation with one comparison star.  The long-standing traditional solution is to use a whole sequence of comparison stars, strung out over a wide range of brightnesses, so that the radius-versus-magnitude relation is empirically determined for the brightness of the target star.

The image diameters can be measured with machines called `iris diaphragm photometers' (first developed in the 1930s) and with photoelectric scanners (first developed in the 1970s).  From the earliest days, the dominant method for measuring image diameters was simply for a trained observer to make size comparisons between the diameter of the target star and the diameters of nearby comparison stars.  The human eye is remarkably accurate at side-by-side comparisons of the sizes of round objects.  The procedure is to view the glass plate on a light table through a loupe or a low-power microscope, compare the target's size to the size of nearby stars, and judge by-eye the relative placement of the target star's diameter.  To give a specific and typical example, if the target is judged to be halfway between two comparison stars of magnitude 12.2 and 12.6, then the target has a magnitude of 12.4.  In practice, an inexperienced worker can produce magnitudes with an accuracy of 0.3 mag or so, while an experienced worker can produce magnitudes with a one-sigma uncertainty of 0.1 mag.  For many situations, where magnitudes from many plates can be averaged together, the real accuracy of the light curves can be 0.02 mag or better.  For an experienced worker, the by-eye method provides equal or better accuracy as compared to iris diaphragm photometers or scanning.  The by-eye method is very simple, cheap, and fast, whereas the instrumented methods are always complex, slow, and costly.

Harvard College Observatory has a collection of roughly 500,000 glass plates recording the entire sky from 1889 to 1989.  (There are few plates from 1954 to 1969 due to the notorious `Menzel Gap'.)  These are almost all blue-sensitive emulsion on glass plates 8$\times$10 inch in size, stored in paper envelopes, and placed on shelves in time order.  The plates were taken with a wide variety of telescope, from essentially camera lenses up to 24-inch apertures, with the plate sizes covering widths from 5$\degr$ to 42$\degr$.  The limiting magnitudes for the normal-quality plates vary from B=12 to deeper than B=18.  The Harvard plates have 1000-4000 plates covering any given position on the sky.  Harvard has about half the existing archival plates in the world, and is nearly the only source for targets in the southern skies.

Historically, from 1890 to 1960, the Harvard plates dominated the world of variable stars for anything fainter than about eleventh magnitude.  To take an example of the hot RCB stars, all four were discovered with the Harvard plates, and the only published information of any type from before the 1950s is the Harvard light curves.  For many questions of modern astrophysics, light curves with 0.1 mag or 0.02 mag accuracy are more than adequate, so the accuracy attainable with CCDs is completely irrelevant.  In the world of variable stars, the stars are displaying phenomena on all time scales.  Modern studies can cover variable star phenomena on time scales faster than the duration of a single telescope run, and multiple telescope runs can be pasted together to get a picture of phenomena up to a decade in time scale.  But to measure phenomena on time scales from a decade to a century, the only means is to use archival data.  For most stars, the only source of archival information older than a decade or two is from archival photographic plates, and that largely means the Harvard plates.  For the hot RCB stars, in looking for any secular trend (as associated with the evolution of the stars), the only solution is to get fully calibrated light curves from Harvard.

Historically, the Harvard plates were the predecessor of the Johnson B system through the North Polar Sequence.  In modern times, the native magnitudes of the Harvard blue plates have always been found to have a near-zero color term with respect to the Johnson B system.  This means that as long as the comparison star magnitudes are on the Johnson B system, then the resultant magnitudes are exactly in the Johnson B system.

For my measures of the Harvard plates, I have taken all my comparison star magnitudes from the B-band measures of the AAVSO Photometric All-Sky Survey ({\it APASS}, Henden \& Munari 2014).  These magnitudes are tied to the Johnson B magnitudes with high accuracy (Munari et al. 2014), as calibrated from the standard stars of Landolt (2009).  Thus, my modern measures of the Harvard plates are accurately in the Johnson B system.

Critically, both the very extensive {\it DASCH} program (Grindlay et al. 2012) and my own extensive measures prove that long term light curves from Harvard of normal (i.e., constant) stars do {\it not} produce any measurable slope or trend (i.e., typically $<$0.05 magnitude per century) over the last century.  Further, these check star magnitudes are consistent with the modern measures.  This is the proof that any observed secular trend is not some data or analysis artifact.

\section{Century-Long Light Curves For Hot RCB Stars}

The goal of this paper is to get the century-long light curves for all four known hot RCB stars so as to test for any secular fading in the maximum light.  For this, the only way to get the old data is from Harvard, and these are only in the Johnson B-band.  To minimize the mixing of bands, I will take the AAVSO and literature magnitudes for the B-band.

The Johnson B magnitudes for the four Hot RCB stars are listed in Table 1.  These do not include the magnitudes where the star was substantially fainter than the maximum brightness.

\begin{table}
	\centering
	\caption{B Magnitudes from Harvard Plates.}
	\label{tab:table1}
	\begin{tabular}{llll} 
		\hline
		Star & Julian Date & B (mag) & Plate\\
		\hline
DY Cen	&	2415898	&	12.7	&	B29838	\\
DY Cen	&	2416255	&	13.0	&	B31827	\\
DY Cen	&	2416959	&	13.3	&	AM3470	\\
DY Cen	&	2417257	&	12.6	&	AM4107	\\
DY Cen	&	2418405	&	12.8	&	B40009	\\
DY Cen	&	2418428	&	12.6	&	AK286	\\
DY Cen	&	2418437	&	12.6	&	AM6102	\\
DY Cen	&	2418507	&	12.7	&	AM6372	\\
DY Cen	&	2418869	&	13.3	&	B41635	\\
DY Cen	&	2420960	&	12.2	&	AM11620	\\
DY Cen	&	2421024	&	12.2	&	AM11923	\\
DY Cen	&	2421315	&	12.2	&	AM12906	\\
DY Cen	&	2421333	&	12.2	&	AM12992	\\
DY Cen	&	2421333	&	12.7	&	AM13003	\\
DY Cen	&	2421338	&	13.0	&	AM13037	\\
DY Cen	&	2421342	&	12.7	&	AM13047	\\
DY Cen	&	2422073	&	12.8	&	AM14631	\\
DY Cen	&	2422130	&	12.8	&	AM14775	\\
DY Cen	&	2422137	&	12.7	&	AM14806	\\
DY Cen	&	2422162	&	12.5	&	AM14853	\\
DY Cen	&	2422172	&	12.6	&	AM14878	\\
DY Cen	&	2422176	&	12.1	&	AM14889	\\
DY Cen	&	2422436	&	12.5	&	AM15123	\\
DY Cen	&	2422437	&	12.4	&	AM15127	\\
DY Cen	&	2422456	&	12.5	&	AM15165	\\
DY Cen	&	2422483	&	12.1	&	AM15231	\\
DY Cen	&	2422493	&	12.1	&	AM15255	\\
DY Cen	&	2422517	&	12.0	&	AM15292	\\
DY Cen	&	2422544	&	12.8	&	AM15382	\\
DY Cen	&	2423180	&	12.5	&	AM15747	\\
DY Cen	&	2426470	&	12.5	&	RB1688	\\
DY Cen	&	2426480	&	12.4	&	RB1729	\\
DY Cen	&	2426490	&	12.7	&	RB1798	\\
DY Cen	&	2426497	&	11.9	&	RB1821	\\
DY Cen	&	2426531	&	12.0	&	RB1900	\\
DY Cen	&	2426546	&	12.8	&	RB1935	\\
DY Cen	&	2426771	&	12.0	&	RB2507	\\
DY Cen	&	2426843	&	12.0	&	RB2753	\\
DY Cen	&	2426899	&	12.7	&	RB3075	\\
DY Cen	&	2431904	&	12.0	&	RB14293	\\
DY Cen	&	2431950	&	12.7	&	RB14385	\\
DY Cen	&	2432011	&	12.3	&	RB14552	\\
DY Cen	&	2432328	&	12.5	&	RB15102	\\
DY Cen	&	2432648	&	12.5	&	RB15596	\\
DY Cen	&	2432681	&	11.9	&	RB15654	\\
DY Cen	&	2432758	&	12.5	&	RB15795	\\
DY Cen	&	2433054	&	12.5	&	RB16284	\\
DY Cen	&	2445490	&	13.8	&	DSB1047	\\
DY Cen	&	2445813	&	13.5	&	DSB1286	\\
DY Cen	&	2445848	&	13.5	&	DSB1325	\\
DY Cen	&	2445872	&	13.1	&	DSB1359	\\
DY Cen	&	2445900	&	13.3	&	DSB1388	\\
DY Cen	&	2446243	&	13.5	&	DSB1708	\\
DY Cen	&	2446257	&	13.6	&	DSB1711	\\
DY Cen	&	2446291	&	13.5	&	DSB1736	\\
DY Cen	&	2446497	&	13.5	&	DSB1903	\\
DY Cen	&	2446527	&	13.2	&	DSB1932	\\
DY Cen	&	2446827	&	13.7	&	DSB2153	\\
DY Cen	&	2446945	&	13.5	&	DSB2227	\\
DY Cen	&	2447002	&	13.8	&	DSB2300	\\
DY Cen	&	2447022	&	13.8	&	DSB2334	\\
DY Cen	&	2447241	&	13.4	&	DSB2493	\\
DY Cen	&	2447267	&	13.3	&	DSB2541	\\
		\hline
	\end{tabular}
\end{table}

\begin{table}
	\centering
	\contcaption{B Magnitudes from Harvard Plates.}
	\label{tab:continued}
	\begin{tabular}{llll} 
		\hline
		Star & Julian Date & B (mag) & Plate\\
		\hline

DY Cen	&	2447298	&	13.7	&	DSB2574	\\
DY Cen	&	2447322	&	13.5	&	DSB2588	\\
DY Cen	&	2447357	&	13.3	&	DSB2630	\\
DY Cen	&	2447380	&	13.6	&	DSB2653	\\
DY Cen	&	2447590	&	13.5	&	DSB2794	\\
DY Cen	&	2447682	&	13.6	&	DSB2827	\\
DY Cen	&	2447761	&	13.6	&	DSB2863	\\
MV Sgr	&	2417077	&	12.6	&	AM3801	\\
MV Sgr	&	2422605	&	13.3	&	MC16930	\\
MV Sgr	&	2422606	&	13.0	&	MC16931	\\
MV Sgr	&	2425746	&	12.7	&	RB334	\\
MV Sgr	&	2425751	&	12.7	&	RB345	\\
MV Sgr	&	2425778	&	12.7	&	RB402	\\
MV Sgr	&	2428306	&	13.0	&	MA5360	\\
MV Sgr	&	2428672	&	12.9	&	MA6375	\\
MV Sgr	&	2429080	&	12.7	&	MA7302	\\
MV Sgr	&	2429429	&	12.1	&	RB8838	\\
MV Sgr	&	2429435	&	12.5	&	RB8861	\\
MV Sgr	&	2429441	&	12.6	&	RB8891	\\
MV Sgr	&	2429485	&	12.3	&	RB9039	\\
MV Sgr	&	2429547	&	12.0	&	RB9181	\\
MV Sgr	&	2429732	&	12.4	&	RB9488	\\
MV Sgr	&	2429793	&	12.4	&	RB9691	\\
MV Sgr	&	2429808	&	12.5	&	RB9732	\\
MV Sgr	&	2429811	&	12.4	&	RB9755	\\
MV Sgr	&	2429869	&	12.4	&	RB9967	\\
MV Sgr	&	2443616	&	13.5	&	DSB512	\\
MV Sgr	&	2444165	&	13.5	&	DSB644	\\
MV Sgr	&	2444821	&	13.5	&	DSB769	\\
MV Sgr	&	2445140	&	13.3	&	DSB911	\\
MV Sgr	&	2445173	&	13.3	&	DSB919	\\
MV Sgr	&	2445551	&	13.2	&	DSB1087	\\
MV Sgr	&	2445801	&	13.5	&	DSB1274	\\
MV Sgr	&	2445824	&	13.5	&	DSB1305	\\
MV Sgr	&	2445858	&	13.0	&	DSB1346	\\
MV Sgr	&	2445908	&	13.5	&	DSB1408	\\
MV Sgr	&	2446210	&	13.6	&	DSB1672	\\
MV Sgr	&	2446233	&	13.3	&	DSB1702	\\
MV Sgr	&	2446294	&	13.3	&	DSB1755	\\
MV Sgr	&	2446624	&	13.0	&	DSB2022	\\
V348 Sgr	&	2413724	&	12.1	&	A1837	\\
V348 Sgr	&	2415533	&	12.0	&	AM808	\\
V348 Sgr	&	2415576	&	11.7	&	AM907	\\
V348 Sgr	&	2415633	&	11.6	&	AM1028	\\
V348 Sgr	&	2415635	&	11.6	&	AM1043	\\
V348 Sgr	&	2417704	&	11.9	&	AM4804	\\
V348 Sgr	&	2417748	&	11.7	&	AM4931	\\
V348 Sgr	&	2417759	&	12.0	&	AM4954	\\
V348 Sgr	&	2417788	&	11.6	&	AM5024	\\
V348 Sgr	&	2417814	&	11.8	&	AM5090	\\
V348 Sgr	&	2417821	&	11.8	&	AM5114	\\
V348 Sgr	&	2418028	&	11.8	&	AM5340	\\
V348 Sgr	&	2418043	&	11.8	&	AM5390	\\
V348 Sgr	&	2418070	&	11.6	&	AM5444	\\
V348 Sgr	&	2418396	&	11.7	&	AM6011	\\
V348 Sgr	&	2418429	&	11.4	&	AK288	\\
V348 Sgr	&	2418439	&	11.6	&	AM6114	\\
V348 Sgr	&	2418454	&	11.7	&	AM6170	\\
V348 Sgr	&	2418502	&	11.8	&	AM6347	\\
V348 Sgr	&	2418532	&	11.7	&	AM6454	\\
V348 Sgr	&	2418822	&	11.8	&	AM6952	\\
V348 Sgr	&	2418849	&	11.8	&	AM7033	\\
V348 Sgr	&	2418856	&	11.6	&	AM7063	\\
		\hline
	\end{tabular}
\end{table}

\begin{table}
	\centering
	\contcaption{B Magnitudes from Harvard Plates.}
	\label{tab:continued}
	\begin{tabular}{llll} 
		\hline
		Star & Julian Date & B (mag) & Plate\\
		\hline

V348 Sgr	&	2419205	&	11.4	&	AM7457	\\
V348 Sgr	&	2419205	&	11.5	&	AM7458	\\
V348 Sgr	&	2419234	&	11.9	&	AM7549	\\
V348 Sgr	&	2419562	&	12.0	&	AM8301	\\
V348 Sgr	&	2419563	&	12.0	&	AM8307	\\
V348 Sgr	&	2419594	&	11.8	&	AM8423	\\
V348 Sgr	&	2419605	&	12.1	&	AM8492	\\
V348 Sgr	&	2419618	&	11.9	&	AM8522	\\
V348 Sgr	&	2419633	&	11.6	&	AM8559	\\
V348 Sgr	&	2422084	&	12.1	&	AM14682	\\
V348 Sgr	&	2422152	&	12.1	&	MC16930	\\
V348 Sgr	&	2422152	&	12.0	&	MC16931	\\
V348 Sgr	&	2422515	&	12.2	&	MC16838	\\
V348 Sgr	&	2422517	&	11.8	&	AM15294	\\
V348 Sgr	&	2422581	&	11.9	&	AM15486	\\
V348 Sgr	&	2422582	&	11.5	&	MF07104	\\
V348 Sgr	&	2423179	&	11.8	&	AM15745	\\
V348 Sgr	&	2423182	&	11.6	&	AM15767	\\
V348 Sgr	&	2423192	&	11.9	&	A11979	\\
V348 Sgr	&	2423195	&	11.8	&	AM15795	\\
V348 Sgr	&	2423210	&	11.5	&	AM15842	\\
V348 Sgr	&	2423223	&	11.7	&	AM15869	\\
V348 Sgr	&	2423236	&	11.8	&	AM15909	\\
V348 Sgr	&	2423248	&	11.5	&	AM15931	\\
V348 Sgr	&	2423249	&	11.6	&	AM15934	\\
V348 Sgr	&	2423347	&	11.9	&	AM16120	\\
V348 Sgr	&	2423663	&	11.5	&	AM16382	\\
V348 Sgr	&	2425706	&	12.0	&	RB228	\\
V348 Sgr	&	2425746	&	12.1	&	RB334	\\
V348 Sgr	&	2425751	&	11.9	&	RB345	\\
V348 Sgr	&	2425778	&	11.9	&	RB402	\\
V348 Sgr	&	2425798	&	11.9	&	RB434	\\
V348 Sgr	&	2426802	&	11.7	&	RB2554	\\
V348 Sgr	&	2426810	&	11.9	&	RB2611	\\
V348 Sgr	&	2426871	&	11.8	&	RB2851	\\
V348 Sgr	&	2426872	&	11.8	&	RB2869	\\
V348 Sgr	&	2427901	&	12.1	&	RB6045	\\
V348 Sgr	&	2428013	&	11.7	&	RB6313	\\
V348 Sgr	&	2428035	&	11.9	&	AM17031	\\
V348 Sgr	&	2428041	&	11.6	&	AM17056	\\
V348 Sgr	&	2429485	&	11.9	&	RB9039	\\
V348 Sgr	&	2429732	&	12.0	&	RB9488	\\
V348 Sgr	&	2430095	&	11.8	&	RB10453	\\
V348 Sgr	&	2430107	&	11.9	&	AM21531	\\
V348 Sgr	&	2430110	&	12.1	&	AM21549	\\
V348 Sgr	&	2430111	&	11.8	&	RB10532	\\
V348 Sgr	&	2430111	&	12.0	&	RB10534	\\
V348 Sgr	&	2430111	&	11.8	&	RB10535	\\
V348 Sgr	&	2430111	&	11.7	&	RB10540	\\
V348 Sgr	&	2430113	&	11.9	&	RB10560	\\
V348 Sgr	&	2430113	&	11.8	&	RB10566	\\
V348 Sgr	&	2430113	&	11.6	&	RB10568	\\
V348 Sgr	&	2430118	&	11.8	&	RB10585	\\
V348 Sgr	&	2430120	&	11.8	&	RB10593	\\
V348 Sgr	&	2430121	&	12.1	&	RB10598	\\
V348 Sgr	&	2430136	&	11.9	&	RB10666	\\
V348 Sgr	&	2430137	&	11.5	&	RB10668	\\
V348 Sgr	&	2430139	&	11.7	&	RB10680	\\
V348 Sgr	&	2430140	&	11.9	&	RB10692	\\
V348 Sgr	&	2430141	&	11.4	&	RB10698	\\
V348 Sgr	&	2430141	&	11.6	&	RB10699	\\
V348 Sgr	&	2430153	&	12.0	&	RB10768	\\
V348 Sgr	&	2430162	&	11.6	&	RB10827	\\
V348 Sgr	&	2430163	&	11.8	&	RB10828	\\
		\hline
	\end{tabular}
\end{table}

\begin{table}
	\centering
	\contcaption{B Magnitudes from Harvard Plates.}
	\label{tab:continued}
	\begin{tabular}{llll} 
		\hline
		Star & Julian Date & B (mag) & Plate\\
		\hline

V348 Sgr	&	2430163	&	11.7	&	RB10829	\\
V348 Sgr	&	2430220	&	11.7	&	AM21975	\\
V348 Sgr	&	2430221	&	11.9	&	AX4088	\\
V348 Sgr	&	2430299	&	11.9	&	RB11123	\\
V348 Sgr	&	2431158	&	12.1	&	RB12537	\\
V348 Sgr	&	2431172	&	11.6	&	AM23552	\\
HV 2671	&	2413878	&	16.2	&	A2172	\\
HV 2671	&	2414253	&	15.4	&	B20843	\\
HV 2671	&	2416398	&	15.4	&	B32728	\\
HV 2671	&	2416817	&	16.0	&	A7098	\\
HV 2671	&	2423466	&	16.3	&	A12286	\\
HV 2671	&	2423487	&	16.1	&	A12288	\\
HV 2671	&	2423683	&	16.2	&	A12699	\\
HV 2671	&	2423684	&	15.8	&	A12700	\\
HV 2671	&	2423707	&	16.1	&	A12788	\\
HV 2671	&	2423733	&	16.0	&	A12830	\\
HV 2671	&	2423735	&	16.1	&	A12834	\\
HV 2671	&	2423738	&	16.1	&	A12848	\\
HV 2671	&	2423739	&	16.2	&	A12851	\\
HV 2671	&	2423741	&	15.5	&	A12855	\\
HV 2671	&	2423753	&	16.2	&	A12865	\\
HV 2671	&	2425941	&	16.3	&	A14366	\\
HV 2671	&	2426309	&	16.0	&	A15041	\\
HV 2671	&	2426309	&	15.5	&	MF15038	\\
HV 2671	&	2426322	&	16.3	&	A15064	\\
HV 2671	&	2426328	&	15.9	&	A15075	\\
HV 2671	&	2426335	&	16.1	&	A15087	\\
HV 2671	&	2426410	&	16.1	&	A15233	\\
HV 2671	&	2426412	&	15.9	&	A15250	\\
HV 2671	&	2426413	&	16.1	&	A15254	\\
HV 2671	&	2426414	&	16.2	&	A15256	\\
HV 2671	&	2426421	&	16.1	&	A15264	\\
HV 2671	&	2426426	&	16.1	&	A15266	\\
HV 2671	&	2426441	&	16.1	&	A15278	\\
HV 2671	&	2426444	&	16.2	&	A15287	\\
HV 2671	&	2426452	&	16.3	&	A15293	\\
HV 2671	&	2426453	&	16.1	&	A15298	\\
HV 2671	&	2426454	&	15.9	&	A15303	\\
HV 2671	&	2426455	&	16.1	&	A15308	\\
HV 2671	&	2426456	&	16.3	&	A15314	\\
HV 2671	&	2426566	&	16.3	&	A15631	\\
HV 2671	&	2426568	&	16.4	&	A15651	\\
HV 2671	&	2426569	&	16.3	&	A15661	\\
HV 2671	&	2426572	&	16.4	&	A15680	\\
HV 2671	&	2426573	&	16.2	&	A15686	\\
HV 2671	&	2426578	&	16.4	&	A15703	\\
HV 2671	&	2426606	&	15.2	&	MF16077	\\
HV 2671	&	2426608	&	15.3	&	MF16082	\\
HV 2671	&	2426636	&	16.4	&	A15806	\\
HV 2671	&	2426657	&	15.7	&	MF16170	\\
HV 2671	&	2426679	&	16.4	&	A15838	\\
HV 2671	&	2426680	&	15.6	&	MF16250	\\
HV 2671	&	2426684	&	16.3	&	A15847	\\
HV 2671	&	2426687	&	16.0	&	A15851	\\
HV 2671	&	2426687	&	15.7	&	MF16282	\\
HV 2671	&	2426690	&	16.2	&	A15858	\\
HV 2671	&	2426710	&	16.3	&	A15872	\\
HV 2671	&	2426710	&	15.5	&	MF16324	\\
HV 2671	&	2426720	&	15.4	&	MF16389	\\
HV 2671	&	2426802	&	15.6	&	MF16591	\\
HV 2671	&	2426931	&	15.3	&	B56513	\\
HV 2671	&	2426946	&	15.6	&	B56559	\\
HV 2671	&	2426947	&	16.2	&	A16203	\\
HV 2671	&	2426950	&	15.7	&	B56593	\\
		\hline
	\end{tabular}
\end{table}

\begin{table}
	\centering
	\contcaption{B Magnitudes from Harvard Plates.}
	\label{tab:continued}
	\begin{tabular}{llll} 
		\hline
		Star & Julian Date & B (mag) & Plate\\
		\hline

HV 2671	&	2426956	&	16.2	&	B56627	\\
HV 2671	&	2426957	&	15.7	&	B56637	\\
HV 2671	&	2426978	&	16.2	&	A16254	\\
HV 2671	&	2427311	&	15.9	&	A16561	\\
HV 2671	&	2427749	&	16.3	&	A17232	\\
HV 2671	&	2427777	&	16.0	&	A17258	\\
HV 2671	&	2427800	&	16.4	&	A17287	\\
HV 2671	&	2427800	&	16.3	&	A17288	\\
HV 2671	&	2427800	&	16.2	&	A17289	\\
HV 2671	&	2427800	&	16.2	&	A17290	\\
HV 2671	&	2427800	&	16.3	&	A17291	\\
HV 2671	&	2427801	&	16.3	&	A17295	\\
HV 2671	&	2427802	&	15.8	&	A17298	\\
HV 2671	&	2427807	&	16.2	&	A17302	\\
HV 2671	&	2427807	&	16.2	&	A17303	\\
HV 2671	&	2427808	&	16.1	&	A17307	\\
HV 2671	&	2427808	&	16.4	&	A17308	\\
HV 2671	&	2427808	&	16.4	&	A17309	\\
HV 2671	&	2427808	&	16.0	&	A17311	\\
HV 2671	&	2427811	&	16.2	&	A17315	\\
HV 2671	&	2429584	&	16.5	&	A21491	\\
HV 2671	&	2429606	&	16.2	&	B65009	\\
HV 2671	&	2429671	&	16.1	&	B65083	\\
HV 2671	&	2429674	&	16.3	&	A21606	\\
HV 2671	&	2429690	&	16.2	&	A21621	\\
HV 2671	&	2429879	&	15.9	&	B65919	\\
HV 2671	&	2429905	&	16.0	&	A22207	\\
HV 2671	&	2429934	&	16.3	&	A22269	\\
HV 2671	&	2429939	&	16.1	&	A22277	\\
HV 2671	&	2429956	&	16.2	&	A22305	\\
HV 2671	&	2429970	&	16.1	&	A22330	\\
HV 2671	&	2429994	&	16.0	&	A22340	\\
HV 2671	&	2430023	&	16.0	&	B66141	\\
HV 2671	&	2430045	&	15.7	&	MF28653	\\
HV 2671	&	2430057	&	16.2	&	MF22404	\\
HV 2671	&	2430058	&	16.2	&	A22409	\\
HV 2671	&	2430080	&	15.9	&	B66300	\\
HV 2671	&	2430101	&	15.7	&	MF28870	\\
HV 2671	&	2430110	&	15.8	&	MF28953	\\
HV 2671	&	2430111	&	16.3	&	MF28967	\\
HV 2671	&	2430112	&	16.2	&	MF28976	\\
HV 2671	&	2430240	&	16.1	&	B67078	\\
HV 2671	&	2430264	&	15.7	&	B67149	\\
HV 2671	&	2430314	&	16.2	&	A22980	\\
HV 2671	&	2430314	&	16.0	&	B67253	\\
HV 2671	&	2430315	&	16.3	&	A22987	\\
HV 2671	&	2430318	&	16.0	&	A22992	\\
HV 2671	&	2430318	&	16.2	&	A22994	\\
HV 2671	&	2430319	&	16.1	&	A22995	\\
HV 2671	&	2430320	&	16.3	&	A23002	\\
HV 2671	&	2430322	&	16.1	&	A23007	\\
HV 2671	&	2430323	&	16.1	&	A23008	\\
HV 2671	&	2430324	&	16.0	&	A23011	\\
HV 2671	&	2430325	&	16.0	&	A23017	\\
HV 2671	&	2430328	&	16.1	&	A23020	\\
HV 2671	&	2430372	&	16.0	&	B67322	\\
HV 2671	&	2430373	&	16.2	&	A23044	\\
HV 2671	&	2430373	&	15.9	&	B67325	\\
HV 2671	&	2430373	&	16.0	&	B67327	\\
HV 2671	&	2430373	&	16.2	&	A23046	\\
HV 2671	&	2430373	&	16.0	&	A23047	\\
HV 2671	&	2430375	&	15.8	&	B67328	\\
HV 2671	&	2430375	&	15.7	&	B67330	\\
HV 2671	&	2430586	&	16.1	&	A23415	\\
		\hline
	\end{tabular}
\end{table}

\begin{table}
	\centering
	\contcaption{B Magnitudes from Harvard Plates.}
	\label{tab:continued}
	\begin{tabular}{llll} 
		\hline
		Star & Julian Date & B (mag) & Plate\\
		\hline

HV 2671	&	2430591	&	16.1	&	A23424	\\
HV 2671	&	2430591	&	15.9	&	B67968	\\
HV 2671	&	2430594	&	16.0	&	A23427	\\
HV 2671	&	2430606	&	16.2	&	A23430	\\
HV 2671	&	2430621	&	16.1	&	A23450	\\
HV 2671	&	2430621	&	15.8	&	B68040	\\
HV 2671	&	2430625	&	16.1	&	A23453	\\
HV 2671	&	2430640	&	16.3	&	A23458	\\
HV 2671	&	2430641	&	16.3	&	A23462	\\
HV 2671	&	2430642	&	16.0	&	A23466	\\
HV 2671	&	2430648	&	16.1	&	A23471	\\
HV 2671	&	2430666	&	16.1	&	A23485	\\
HV 2671	&	2430673	&	16.2	&	A23490	\\
HV 2671	&	2430696	&	16.3	&	A23502	\\
HV 2671	&	2430713	&	16.2	&	A23513	\\
HV 2671	&	2430749	&	15.9	&	A23528	\\
HV 2671	&	2430750	&	16.2	&	A23530	\\
HV 2671	&	2430766	&	15.4	&	MF31282	\\
HV 2671	&	2430767	&	16.2	&	A23570	\\
HV 2671	&	2430782	&	15.6	&	MF31352	\\
HV 2671	&	2430791	&	15.1	&	MF31364	\\
HV 2671	&	2430792	&	15.3	&	MF31381	\\
HV 2671	&	2430793	&	15.5	&	MF31390	\\
HV 2671	&	2430809	&	15.7	&	B68351	\\
HV 2671	&	2431804	&	16.3	&	A25189	\\
HV 2671	&	2431814	&	15.5	&	B71365	\\
HV 2671	&	2431817	&	15.8	&	MF35012	\\
HV 2671	&	2431823	&	16.1	&	A25194	\\
HV 2671	&	2431873	&	16.2	&	A25218	\\
HV 2671	&	2431874	&	16.0	&	B71427	\\
HV 2671	&	2432070	&	16.3	&	B72205	\\
HV 2671	&	2432070	&	16.3	&	A25565	\\
HV 2671	&	2432940	&	15.6	&	A26696	\\
HV 2671	&	2433161	&	15.9	&	A26976	\\
HV 2671	&	2433181	&	16.2	&	A26998	\\
		\hline
	\end{tabular}
\end{table}

The archived magnitudes from the AAVSO are freely available on-line.  The AAVSO B-band measures are all taken with CCDs, with photometric uncertainties of 0.03 mag or better.  Observers are designated with a three-letter designation, with HMB being Dr. Franz-Josef Hambsch in Belgium, DSI being Giorgio di Scala in Australia, and SXN being Michael Simonson in the United States.  These are all calibrated with APASS comparison stars, and are thus in the Johnson B system.  

I have also pulled a variety of magnitudes from the literature, and these are all CCD measures.  (The one exception is the single magnitude from Herbig in 1964 for MV Sgr.)  These have been calibrated ultimately from the Landolt fields, and thus are also in the Johnson B system.  Intercomparison of modern published B magnitudes always shows that different sources disagree with each other up to $\sim$0.1 mag, even for known-constant stars and for effectively simultaneous measures of slow variable stars.  This is likely being due to different color terms and calibrations between observers.  Within each literature source, the quoted error bars are usually $\sim$0.01 mag, but these are always measurement errors and do not include systematic errors that will appear as a constant offset for each source.  Fortunately, the hot RCB stars show variations that are greatly larger than these usual calibration problems, so the existence and slope of the trends remain unaffected.  In all, to within the normal errors, all the literature magnitudes are on the modern Johnson B system.

The archival magnitudes in the literature (Hoffleit 1930; 1958; 1959; Woods 1928) are not used, because all have big photometric differences from the modern B magnitude system due to problems with the comparison star sequences, as was universal for the era.  I have examined the exact same plates at Harvard, plus many more, all on the modern Johnson B system, so my magnitudes now supersede the old ones in the literature.

\begin{figure}
	\includegraphics[width=1.1\columnwidth]{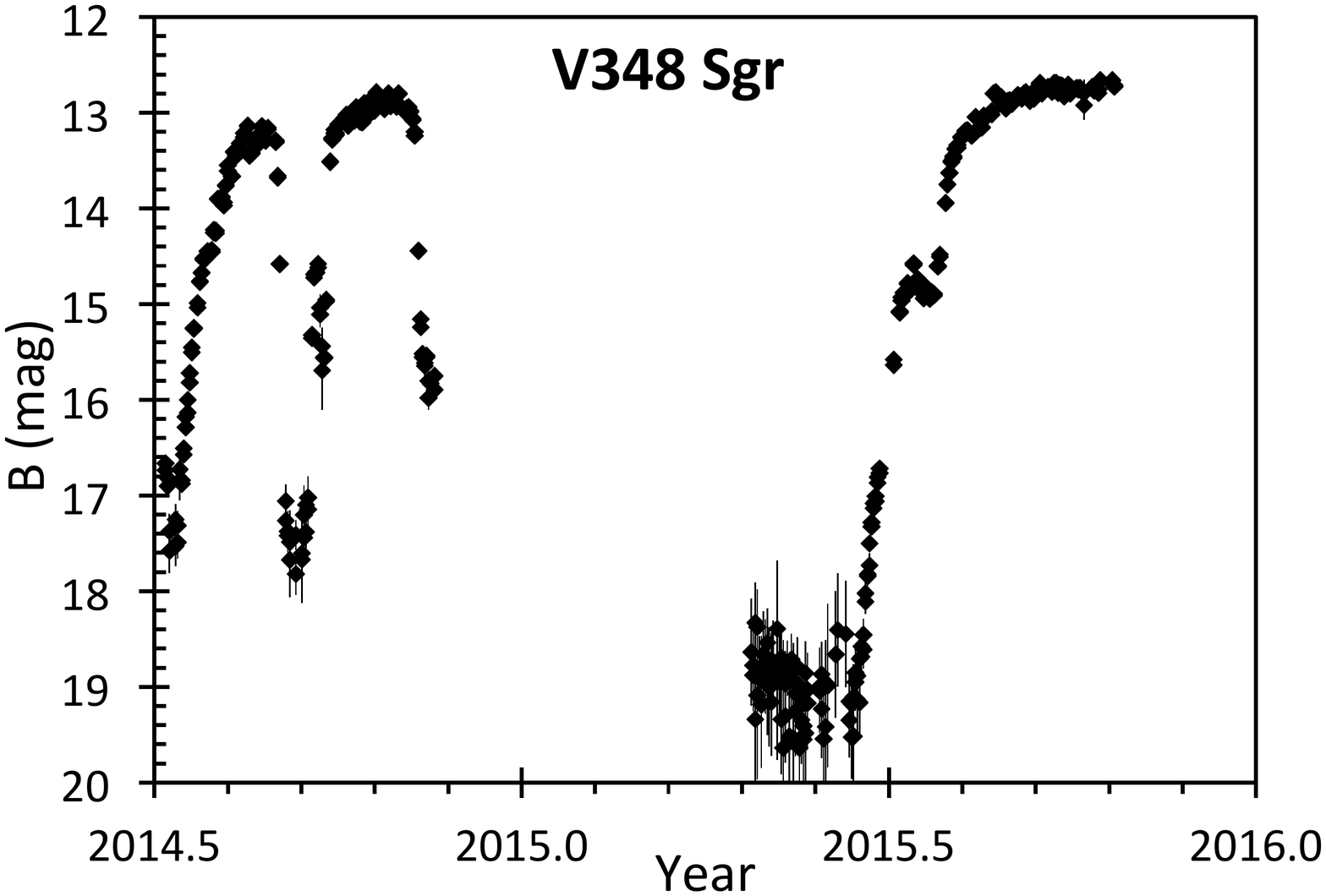}
    \caption{V348 Sgr in B from AAVSO in 2014-2015.  The observer was Dr. Franz-Josef Hambsch, in Belgium with a 14-inch telescope.  This Johnson B light curve has 483 points, for which 164 magnitudes have been selected as representing the star at maximum light, while Hambsch also has Johnson V magnitudes on all the same nights.  This light curve illustrates that the complete recovery from a dip only asymptotically approaches some presumably-dust-free maximum.  It also illustrates that the time duration when the star is in the dip but just below maximum is a very small fraction of the time.  A further point is that we can confidently measure the magnitudes at maximum to better accuracy than the maximum can be defined.  The main point of this figure is that the recent maximum of V348 Sgr is close to B=12.93, whereas the Harvard plates show a maximum light around B=11.8 over a century earlier, with this being proof of a secular decline.}
    \label{fig:Fig. 1}
\end{figure}

A substantial problem is to select out the magnitudes taken when the star is at maximum light.  Part of the problem is that the brightness recovery from a decline is only asymptotic, so all magnitudes will still have some residual dust to varying degrees.  This is illustrated in Fig. 1, where V348 Sgr never quite recovers completely to some dust-free maximum brightness.  An adequate solution is simply that this effect should be the same for old and new magnitudes, so there should be negligible effect on any trends.  The biggest part of the problem is that most of the magnitudes are isolated in time, so we cannot recognize whether the star is at maximum or is in a dip.  Magnitudes greatly fainter than some maximum are easily recognized and rejected, but magnitudes from the start or end of a dip, with the brightness only somewhat below the true maximum, can be included, resulting in an apparent fainter maximum.  The inclusion of more or fewer in-decline magnitudes will make the star's maximum appear to be fainter or brighter.  Fortunately, this problem is minimized by several means.  First, dips are deep with fast drop offs, so there will be only a small fraction of the time during which the star will be close-but-below maximum light.  That is, contaminated magnitudes must be rare and statistically negligible.  Second, for DY Cen and MV Sgr, the dips are rare, so there is little opportunity for contamination.  Third, I have rejected plates taken near times of known dips, whether or not the plate shows the star near a maximum.  Fourth, for the AAVSO light curves, there are a high density of observations so that dips can be easily recognized (e.g., see Fig. 1) and avoided.  Fifth, in generating a light curve at maximum, the effect of including magnitudes in dips) will only matter for measuring secular fading if the early and late measures have different inclusion fractions for dip-magnitudes, and this does not seem plausible.  In general, operationally, when I have no additional information, I have tossed out magnitudes if they are more than a magnitude fainter than the maximum for that star and decade.  There is a plausible chance that inclusion of the initial and final parts of dips has slightly lowered some of the averages over time.  In general, the problem is likely to be minimal in the averages, and certainly the effect is smaller than the trends observed.  Thus, I conclude that this problem is not a significant contributor to the observed trends for any of the hot RCB stars. 

DY Cen has had no minimum from 1960 to 2016, as shown by the densely-sampled light curves from the Royal Astronomical Society of New Zealand and from the AAVSO (De Marco et al. 2002).  With the Harvard plates, I can extend this back to 1935, although the interval from 1954 to 1960 is poorly covered due to the Menzel Gap.  Before 1930, Hoffleit (1930) identified four dust dips with the Harvard plates, while I have added further dips.  The known dips are in 1897, 1901, 1904.4, 1906.3-1908.5, 1914.5, 1915.3, 1918.5, 1924.1-1924.6, 1929.2-1929.5, 1931.2, 1932.6 and 1934.2.  The coverage of the dips is patchy, but it appears that durations are a few months, other than the cases noted.  Further dips are likely to have occurred, mainly during the part of the year when DY Cen is the closest to the Sun.  We are left with a stark situation where DY Cen has many dust dips from 1895 to 1934, but none from 1935 to 2016.

With this, I have constructed a maximum light curve for each of the hot RCB stars with Harvard, AAVSO, and literature magnitudes, all in the Johnson B system.  A simple plot of all these magnitudes shows the usual scatter, with this somewhat hiding secular trends.  To solve this, I have averaged the magnitudes by source and time interval.  The one-sigma uncertainty is taken to be the RMS scatter in the magnitudes divided by the square root of the number of magnitudes.  These averages are presented in Table 2 and Fig. 2.

We see that all four hot RCB stars have an obvious secular decline of roughly one magnitude per century.  This then provides the direct confirmation of the result in De Marco et al. (2002).  The light curves show roughly linear declines.  There is substantial scatter around these linear declines, with it being unclear whether this is due to real variation in the star's maximum light, due to inclusion of magnitudes just below maximum, or due to measurement error.  The secular decline need not be linear or even monotonic.  We can quantify the secular decline by an average decline rate, derived from a linear fit.  

I have made chi-square fits for the light curves in Table 1 for a linear decline.  The resultant fits have reduced chi-square values that are much greater than unity, pointing to the variations around the simple straight line being much larger than the nominal error bars.  As such, the formal one-sigma error bars from the chi-square fits for the slope are not meaningful.  The fitted slopes are 1.15 for DY Cen, 1.29 for MV Sgr, 1.29 for V348 Sgr, and 0.73 for HV 2671, all in units of magnitudes per century.  For these four slopes, the average is 1.11 magnitude per century, with an RMS of 0.26 magnitude per century.  The calculation of an averaged linear slope is making no implication that any of the stars has a constant linear slope, nor that the stars all have the same linear slope.  Indeed, for DY Cen, the light curve appears to be more of a parabola than a line.  Further, the Hot RCB stars are apparently a diverse class, so an average decline rate will be some sort of a mixture of rates for stars with different histories and masses.  Still, the average fade rate of 1.11$\pm$0.26 magnitude per century has utility in expressing the typical decline rate and its variations, in quantitatively showing that the Hot RCB stars are fast fading, and in providing a representative rate for model calculations.

While still with only one band, the AAVSO visual light curves are long enough and with enough accuracy that we can get an independent measure of the secular fading rate.  For DY Cen, 6438 visual magnitudes cover the time from April 1978 to October 2015 with no dips, with an average decline rate of 1.87 magnitudes per century.  For V348 Sgr, the frequent dips make it harder to pick out a decline by eye from the full visual light curve, yet the maximum magnitudes are around 11.5 in the 1950's and around 12.2 for the last decade, for a decline rate of approximately 1.3 magnitudes per century.

\begin{table}
	\centering
	\caption{Hot RCB star light curves.}
	\label{tab:table1}
	\begin{tabular}{llll} 
		\hline
		Star & Years & $\langle$B$\rangle$ (mag) & Source\\
		\hline
DY Cen	&	1902--1910	&	12.84	$\pm$	0.10	&	HCO (9 plates)	\\
DY Cen	&	1916--1922	&	12.46	$\pm$	0.06	&	HCO (21 plates)	\\
DY Cen	&	1931--1932	&	12.33	$\pm$	0.12	&	HCO (9 plates)	\\
DY Cen	&	1946--1949	&	12.36	$\pm$	0.10	&	HCO (8 plates)	\\
DY Cen	&	1970	&	12.62	$\pm$	0.03	&	Marino \& Walker (1971)	\\
DY Cen	&	1972	&	12.70	$\pm$	0.03	&	Sherwood (1975)$^a$	\\
DY Cen	&	1983--1989	&	13.51	$\pm$	0.04	&	HCO (23 plates)	\\
DY Cen	&	1983	&	12.96	$\pm$	0.02	&	Kilkenny et al. (1985)	\\
DY Cen	&	1985	&	13.03	$\pm$	0.02	&	Goldsmith et al. (1990)	\\
DY Cen	&	1987	&	13.11	$\pm$	0.02	&	Pollacco \& Hill (1991)	\\
DY Cen	&	1988	&	13.22	$\pm$	0.04	&	Jones et al. (1989)	\\
DY Cen	&	2006--2007	&	13.45	$\pm$	0.01	&	AAVSO (DSI, 38 mags)	\\
DY Cen	&	2013--2015	&	13.82	$\pm$	0.09	&	AAVSO (SXN, 25 mags)	\\
MV Sgr	&	1905	&	12.60	$\pm$	0.20	&	HCO (1 plate)	\\
MV Sgr	&	1920	&	13.15	$\pm$	0.20	&	HCO (2 plates)	\\
MV Sgr	&	1929	&	12.70	$\pm$	0.20	&	HCO (3 plates)	\\
MV Sgr	&	1934-1940	&	12.48	$\pm$	0.08	&	HCO (13 plates)	\\
MV Sgr	&	1963	&	12.96	$\pm$	0.10	&	Herbig (1964)	\\
MV Sgr	&	1978--1986	&	13.36	$\pm$	0.05	&	HCO (14 plates)	\\
MV Sgr	&	1985	&	13.62	$\pm$	0.03	&	Goldsmith et al. (1990)	\\
MV Sgr	&	2006--2014	&	13.59	$\pm$	0.01	&	AAVSO (DSI, 76 mags)	\\
MV Sgr	&	2011--2015	&	13.90	$\pm$	0.01	&	AAVSO (SXN, 71 mags)	\\
V348 Sgr	&	1896--1901	&	11.80	$\pm$	0.09	&	HCO (5 plates)	\\
V348 Sgr	&	1907--1912	&	11.75	$\pm$	0.03	&	HCO (27 plates)	\\
V348 Sgr	&	1919--1923	&	11.79	$\pm$	0.04	&	HCO (18 plates)	\\
V348 Sgr	&	1929--1935	&	11.87	$\pm$	0.03	&	HCO (13 plates)	\\
V348 Sgr	&	1939--1944	&	11.81	$\pm$	0.03	&	HCO (30 plates)	\\
V348 Sgr	&	1970	&	12.50	$\pm$	0.10	&	Heck et al. (1985)	\\
V348 Sgr	&	1972--1974	&	12.78	$\pm$	0.28	&	Heck et al. (1985)	\\
V348 Sgr	&	1981	&	12.78	$\pm$	0.01	&	Heck et al. (1985)	\\
V348 Sgr	&	2014--2015	&	12.93	$\pm$	0.02	&	AAVSO (HMB, 164 mags)	\\
HV 2671	&	1896--1904	&	15.76	$\pm$	0.26	&	HCO (4 plates)	\\
HV 2671	&	1923	&	16.05	$\pm$	0.10	&	HCO (11 plates)	\\
HV 2671	&	1929--1935	&	16.07	$\pm$	0.04	&	HCO (63 plates)	\\
HV 2671	&	1939--1943	&	16.02	$\pm$	0.03	&	HCO (68 plates)	\\
HV 2671	&	1945--1949	&	16.02	$\pm$	0.08	&	HCO (11 plates)	\\
HV 2671	&	1993--1999	&	16.75	$\pm$	0.1	&	Alcock et al. (1996)	\\
HV 2671	&	2001--2009	&	16.41	$\pm$	0.1	&	Soszynski et al. (2009)	\\
		\hline
	\end{tabular}
	$^a$As quoted in Rao et al. (1993)
\end{table}

\begin{figure}
	\includegraphics[width=1.1\columnwidth]{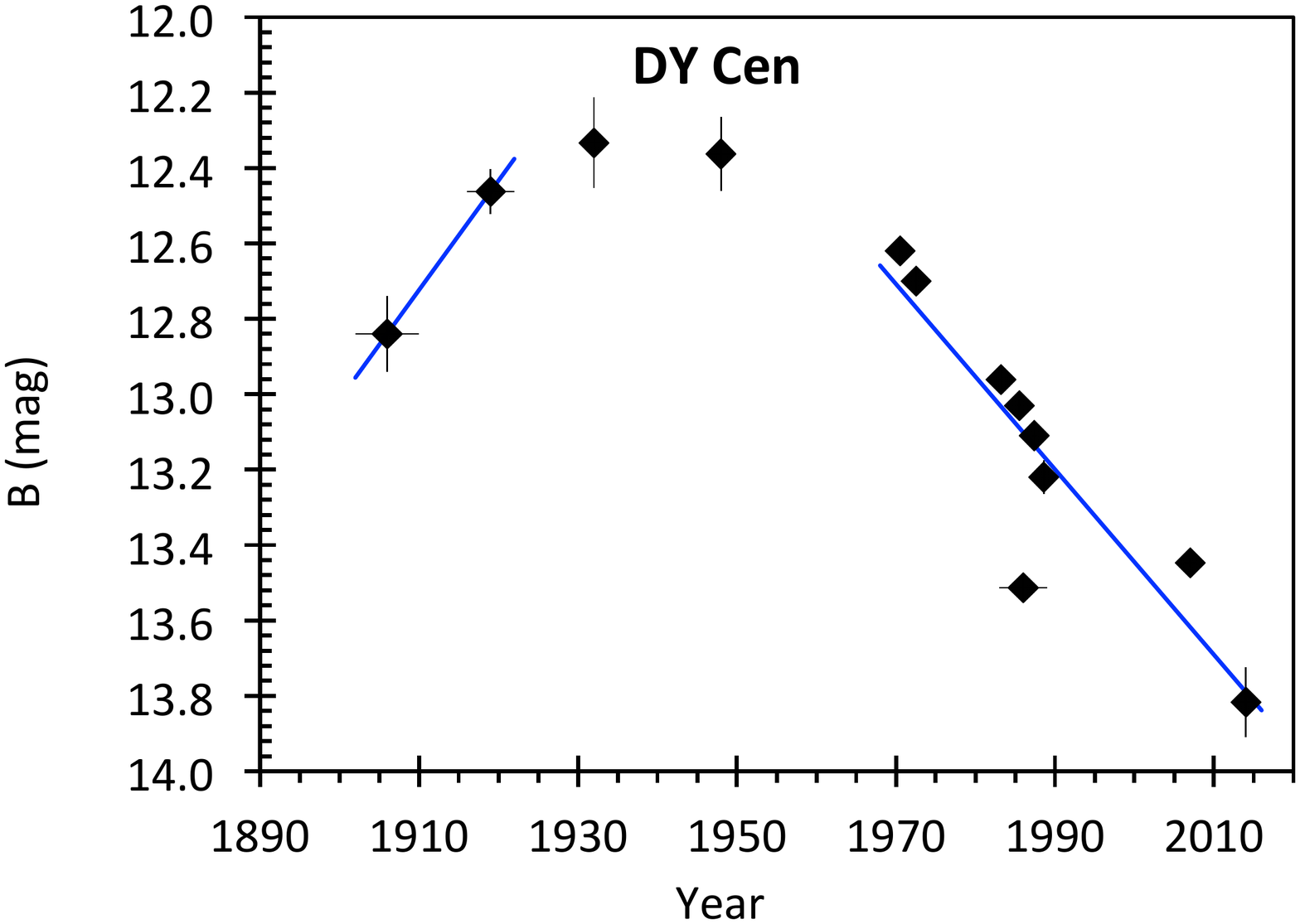}
	\includegraphics[width=1.1\columnwidth]{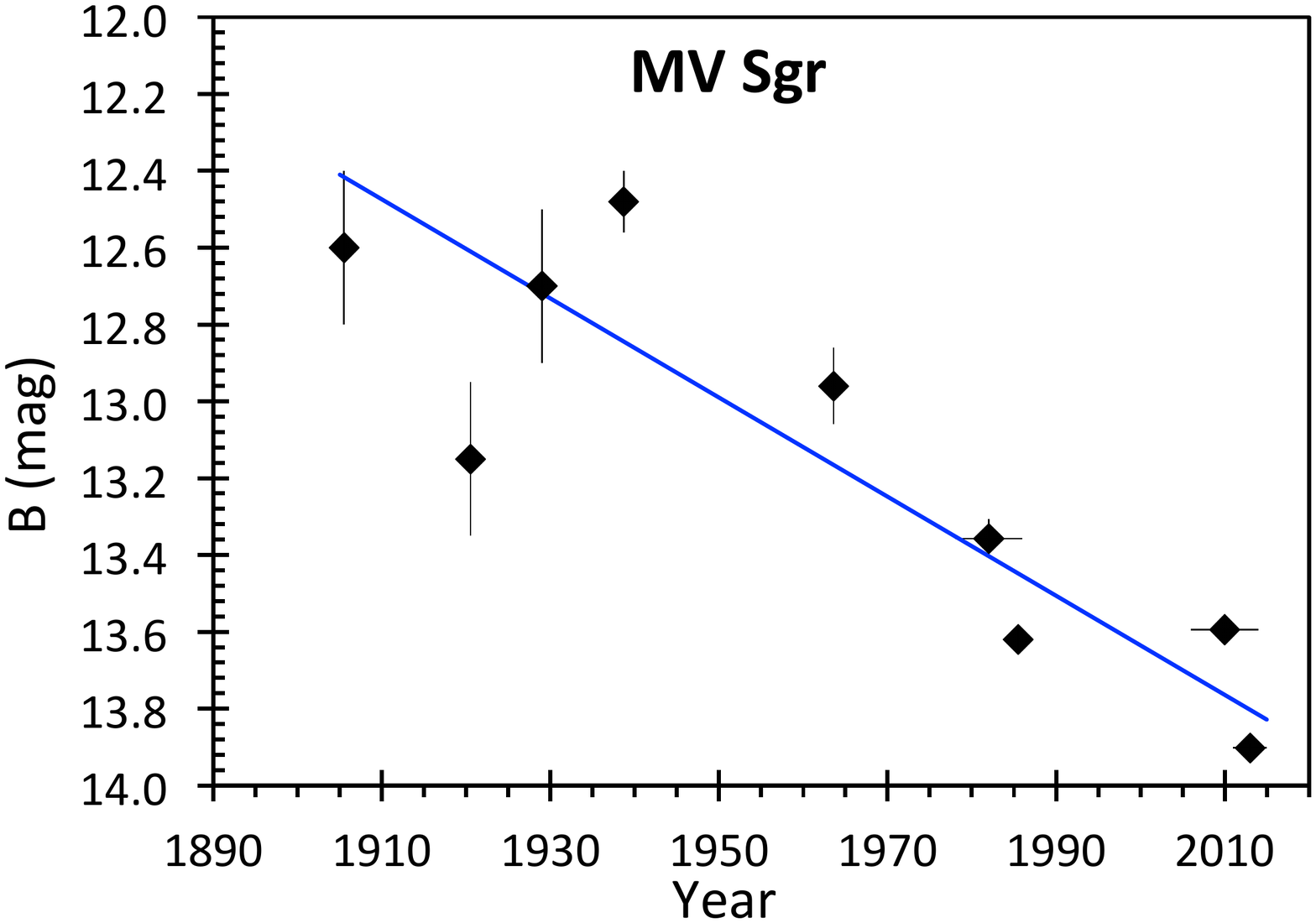}
    \caption{Century-long Johnson B light curves for all four known hot RCB stars.  The main point of this figure and this paper is that all four known hot RCB stars have obvious and significant secular declines.  The thick lines are from the formal chi-square fit, which represent the average secular fading of the stars.  The scatter around these best fits is much larger than the nominal error bars, and it is not clear whether this is due to the intrinsic variations of the maximum brightness, the inclusion of just-below-maximum in-dip magnitudes, or ordinary photometric errors.  The four panels are for DY Cen, MV Sgr, V348 Sgr, and HV 2671.  The fading of DY Cen is apparent only since 1960 or so, whereas the star was {\it brightening} before the 1930s.}
    \label{fig:Fig. 2}
\end{figure}

\begin{figure}
	\includegraphics[width=1.1\columnwidth]{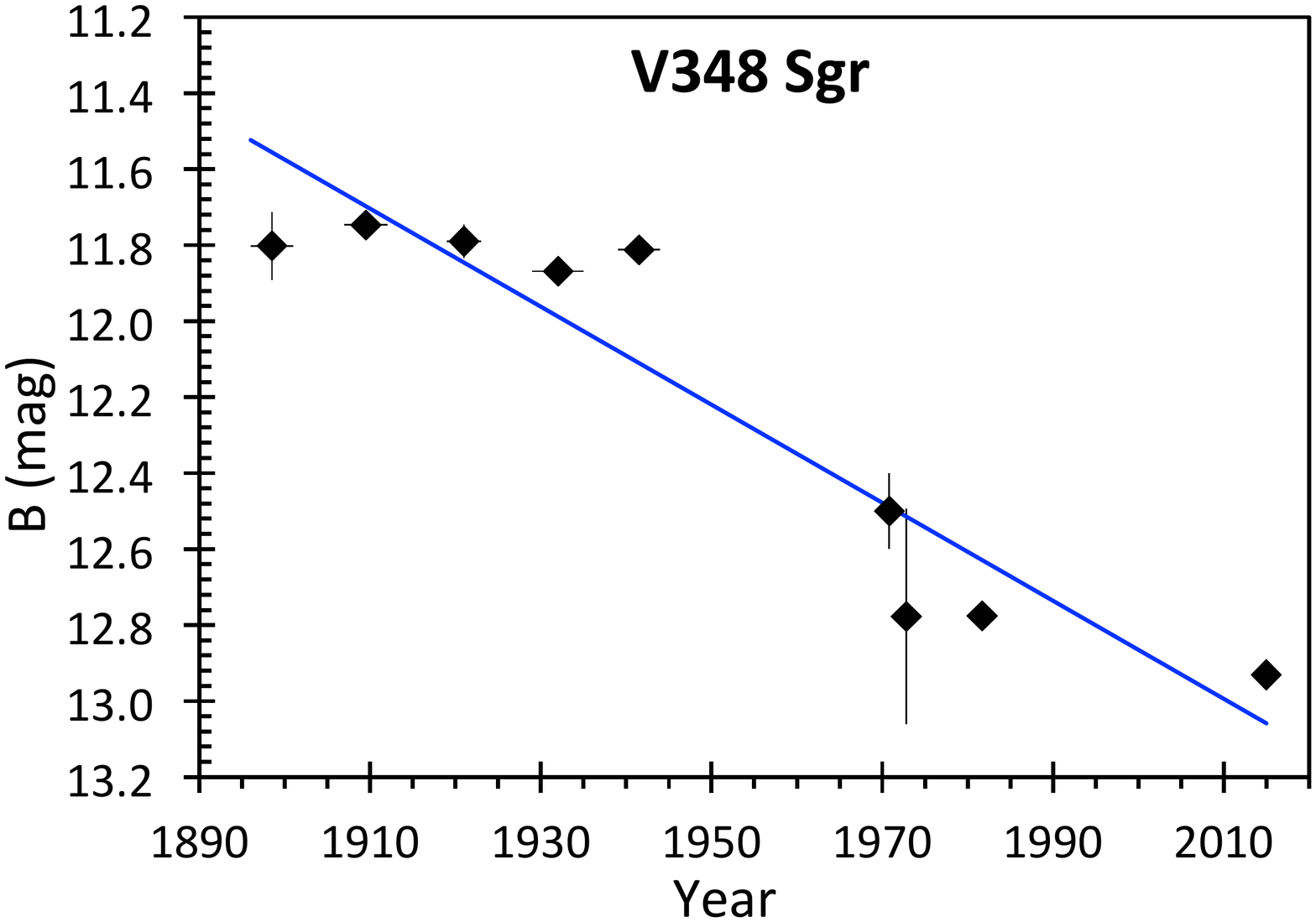}
	\includegraphics[width=1.1\columnwidth]{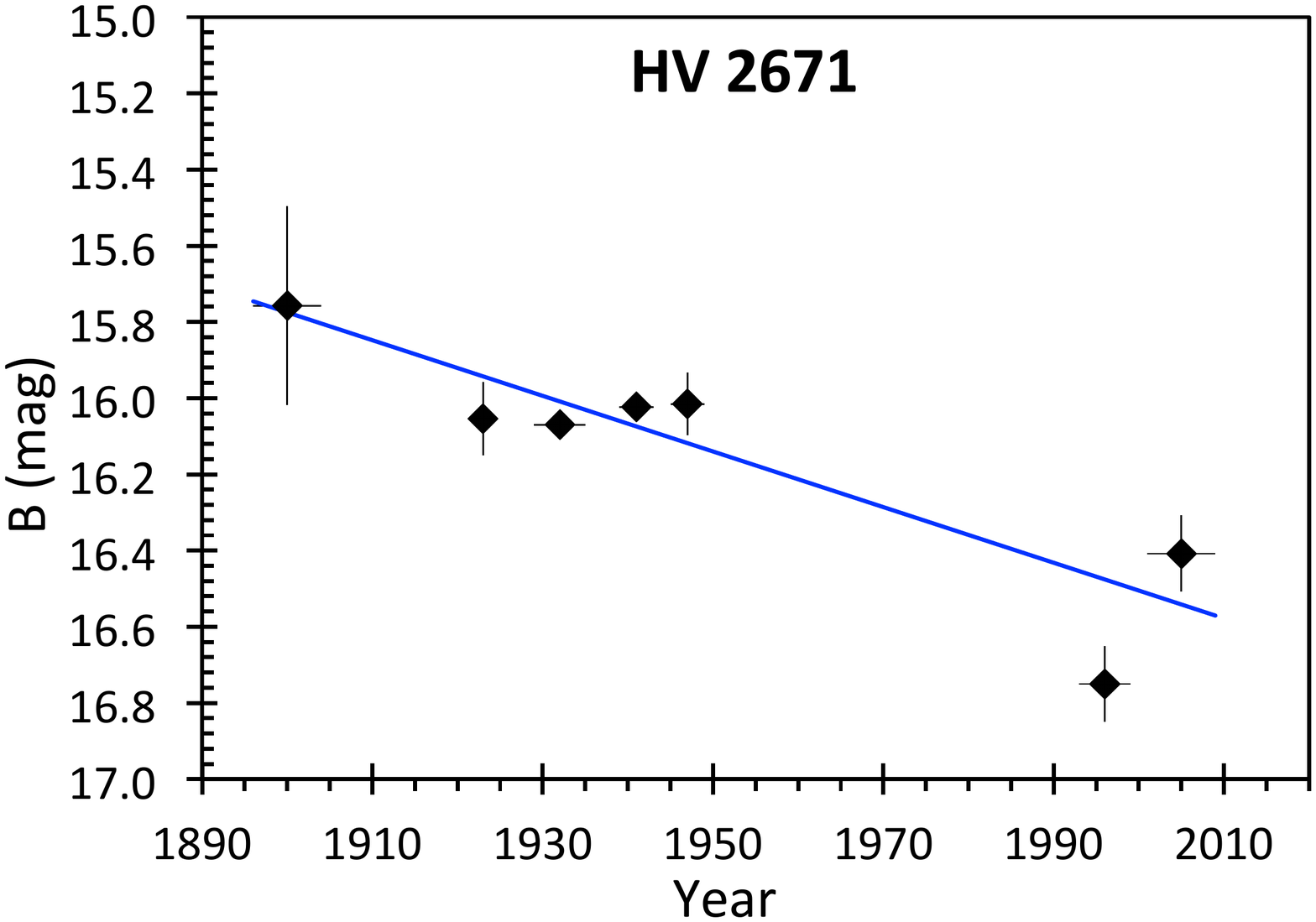}
    \label{fig:Fig. 2cd}
\end{figure}

\section{DIscussion}

In essence, I have merely confirmed and extended the conclusion of De Marco et al. (2002) that the hot RCB stars are secularly fading.  My improvements have been to use a single photometric system all with modern comparison stars, to collect many more magnitudes over a much wider time range, as well as to measure the decline rate for the fourth hot RCB star.

De Marco et al. (2002) have interpreted these secular declines as being due to the star evolving to hotter temperature at constant luminosity, such that the bolometric correction to the optical band makes for an apparent dimming.  (The only other plausible explanation is some sort of general increase in the circumstellar dust density, but such would lead to color changes that are not observed in the cases of DY Cen and MV Sgr.)  This interpretation matches with the general idea that the hot RCB stars are moving horizontally across the top of the HR diagram as part of their normal and fast evolution.  Pandey et al. (2014) have explicitly tested this interpretation for DY Cen, where archival spectra give surface temperatures of 19,400$\pm$400 K in 1987, 23,000$\pm$300 K in 2002, and 24,800$\pm$600 K in 2010.  This is 5,400 K in 23 years, or 23,500 K per century.  This increase in the stellar temperature is confirmed and reflected in the dramatic change in the excitation of the nebula around DY Cen (Rao et al. 2013).

We can translate this rate of temperature change for DY Cen into a magnitude decline rate.  The calculation of the change in bolometric corrections and the change of B-V color is presented in Fig. 1 of Pandey et al. (2014) for the relevant conditions.  For a temperature of 19,400 K, they give V=12.78 and B-V=-0.80 (with an arbitrary zero point), for B=11.98.  For a temperature of 24,800 K, they give V=13.38 and B-V=-0.85 (with the same arbitrary zero point), for B=12.53.  Thus, the observed temperature decline in 23 years corresponds to a fading by 0.55 mag, for a decline rate of 2.39 magnitudes per century.  This is close to the average decline rate for the years 1983 to 2015 (see Fig. 2a).  So the observed temperature change is consistent with the observed decline rate.

For the evolution of DY Cen going back in time, the temperature must be relatively low in old times, resulting in a large bolometric correction.  A simple extrapolation back to 1905 puts the temperature to near zero, so the temperature changes cannot be linear with time.  Nevertheless, the temperature back in 1905 should be relatively quite cold.  The bolometric correction for the B band is minimized for a stellar temperature of 7,500 K, so that for evolution at constant luminosity, the B magnitude will be brightest at that temperature and dimmer as the temperature departs from this value to both hotter and colder temperatures.  So we then have a ready interpretation of the long-term trend in the maximum magnitude (Fig. 2a), with DY Cen starting in 1906 out colder than 7,500 K, heating up to 7,500 K in 1932 when the star was at its brightest, then continuing heating of the star makes it dim over the next decades.  The correction from a constant luminosity to the B band can be taken from Table 15.7 of Cox et al. (2000), where the minimum correction is at 7,500 K, and where the corrections of 0.6 mag are for temperatures of 5,800 K and 11,000 K.  With this, DY Cen had temperatures of 5,800 K around 1906, 7,500 K around 1932, and 11,000 K around the 1970s.   So we now have a simple explanation for why the DY Cen maxima back around 1906 was substantially dimmer than around in 1932.  In all, a continuous temperature increase from 5,800 K in 1905 to 24,800 K in 2010 can account for both the observed change in maximum magnitudes and the observed changes in the temperature.

I have made a crude model that accounts for the stellar temperature and maximum magnitude as a function of time.  From the models of Saio (1988), I take the logarithm of the temperature to be linear with time, with this being approximately right for a given star under hot RCB conditions.  I then set the linear relation by using the observed temperature in 2010 plus the 7,500 K condition for 1932.  With these temperatures, I get the bolometric corrections and B-V colors for supergiants from Cox et al. (2000), and add a constant to get the B-band magnitude outside of a decline.  This model light curve is displayed in Fig. 3.  This model result is not perfect, with the worst problem being that the bolometric correction for 24,800 K should make DY Cen close to 2.0 mag fainter in 2010 than in 1932, whereas it is observed to be more like 1.3 mag fainter.  This problem is easily solved if there is extra light in the DY Cen system, perhaps from a wide binary companion or from the circumstellar material.  Nevertheless, it is clear that the model captures the essence of a normal RCB stars heating up from around 5,800 K in 1906 to 24,800 K in 2010.

\begin{figure}
	\includegraphics[width=1.1\columnwidth]{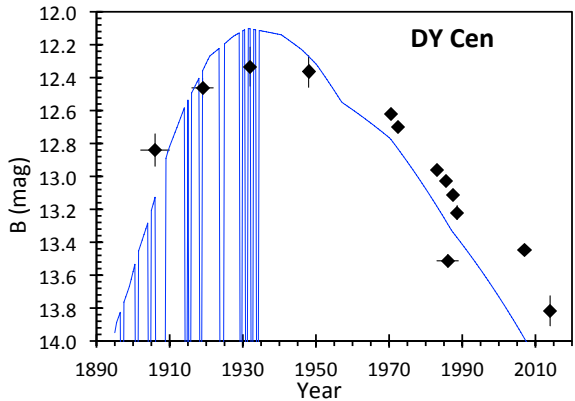}
    \caption{DY Cen evolution from normal RCB to hot RCB to an extreme helium star.  Over the last century, the maximum brightness has first brightened, come to a peak round 1932, then started a secular decline continuing to today.  The temperature is observed to go from 19,400 K in 1987 to 24,800 K in 2010.  This is all consistent with the expected evolution from right to left across the HR diagram at constant luminosity.  DY Cen started in 1906 with a temperature of near 5,800 K as a normal RCB star, and rapidly heated up.  As it heated up, the bolometric correction lessened, making the star appear brighter, until around 1932 when the bolometric correction is its smallest for a temperature near 7,500 K.  As the star kept heating up, the bolometric correction got larger, making the maximum magnitude dim, with this continuing to today.  A crude model of this is shown here, with the logarithm of the temperature assumed to change linearly with time, for comparison with the observed peak magnitude.  The frequency of RCB dips has also change, with many known dips before 1934, but none known after 1934.  All known dips are shown displayed onto the model light curve.}
    \label{fig:Fig. 3}
\end{figure}

So we have actually watched DY Cen start out as an ordinary RCB star (with temperature around 5,800 K), and then heat up to become a hot RCB star, and now appear as an extreme helium star with no dust dips.  This evolution has taken close to one century.  So we have a real measure of the duration of the hot RCB stage, and it is about one century.  This is a very fast phase of evolution.  This explains why so few hot RCB stars have been seen in our Milky Way galaxy, despite them being supergiant stars.  

The heating up of DY Cen is also associated with a sharp drop off in the frequency of dips.  De Marco et al. (2002) note that DY Cen had only four knowns dips from 1897 to 1927, and zero known dips since 1960.  With the Harvard plates, I have identified eight additional dips, all from 1904 to 1934.  Apparently, the heating of the star's surface temperature is connected with the turn off of the dust formation, perhaps caused by a stoppage of pulsations as the star leaves some instability strip.  We realize that there is only a narrow time window over which the hot RCB phenomenon can be recognized, with only a few decades from the time when DY Cen was sufficiently hotter than the upper limit for normal RCB stars up until the time when the dust dips turn off.

We can translate the typical decline rate of 1.11 magnitude per century into a temperature change rate.  For a case with effective surface temperature of 15,000 K, Pandey et al. (2014) give V=12.15 and B-V=-0.75 (with the same arbitrary zero point), for B=11.40.  For a temperature of 20,000 K, they give V=12.88 and B-V=-0.79, for B=12.09.  For a 5,000 K temperature change over the range for hot RCB stars, the B magnitude changes by 0.69 magnitudes.  If this change happens over 62 years, then the B magnitude will fade at the rate of 1.11 magnitude per century. 

DY Cen is similar to extreme helium stars (supergiants composed mostly of helium with near one percent carbon and temperatures 9000-35,000 K).  Jeffery et al. (2001) found that four out of twelve extreme helium stars are heating up with rates from 20 to 120 degrees K per year.  (A useful program would be to search for B-band brightness changes from the 1890s to the present with the Harvard plates for the two stars with the fastest temperature changes; HD 160641 and BD -1$\degr$3438.)  Such surface temperature changes are expected from models of extreme helium stars with masses of $\sim$0.9 M$_{\odot}$ (Saio 1988).  The majority of stars with no measured change in surface temperature are presumably less massive, perhaps $\sim$0.7 M$_{\odot}$.  DY Cen is changing at a rate of 235 K per year from 1987 to 2010.  If the models of Saio (1988) are applicable to DY Cen, then this star would be $\sim$1.0 M$_{\odot}$. 

With the realization as to how some `cold' RCB stars should evolve on a time scale of a century, we can look for the same changes amongst the known normal RCB stars.  That is, the normal RCB stars are heating up, having their maximum magnitudes getting brighter, and their frequency of declines falling to near zero.  But such changes have never been seen for any star that is now a `cold' RCB star.  A small number of cold RCB stars have century-long light curves with no apparent change in their brightness at maximum, while R CrB itself has a 230 year record of unchanging peaks.  So the heating up of the cold RCB stars must usually be too slow to produce observable effects.  Still, some fraction of the now-cold RCB stars might be like DY Cen a century ago.  Perhaps only the most-massive cold RCB stars will be evolving fast enough for the changes to be detectable.

A practical plan to search for fast evolving cold RCB stars is to construct a century-long light curve for many of them.  This could show secular changes in the magnitude at maximum as well as in the frequencies of declines.  In practice, the primary sources are archival data from AAVSO and Harvard.  In any such study, care must be used to place all magnitudes onto a consistent magnitude system.  (For example, old AAVSO magnitudes will require corrections for changes in the comparison sequences, and these can only be gotten from old charts archived at AAVSO Headquarters.)  A substantial problem in seeking changes in the decline-frequency will be to adjust for the variations in time-coverage over the decades.  With this, we have a plan for someone to make a systematic survey of century-long light curves for normal RCB stars so as to measure their evolution across the HR diagram.

In general, stars evolve on such slow time scales that astronomers have not been able to see the changes over time.  Other than for supernovae, evolutionary changes have only been seen for a few post-AGB stars, including the born-again stars and the Stingray (Schaefer \& Edwards 2015).  Now, we can add the four hot RCB stars, with observed temperature rises of 8,000 K or more over the last century.





\begin{thebibliography}{99}
\bibitem[\protect\citeauthoryear{Alcock et al.}{1996}]{Alcock et al. 1996}
Alcock, C., Allsman, R. A., Alves, D. R., et al., 1996, ApJ, 470, 583
\bibitem[\protect\citeauthoryear{Clayton}{1996}]{Clayton 1996}
Clayton, G.~C., 1996, PASP, 108, 225
\bibitem[\protect\citeauthoryear{Clayton}{2012}]{Clayton 2012}
Clayton, G.~C., 2012, JAAVSO, 40, 539
\bibitem[\protect\citeauthoryear{Cox}{2000}]{Cox 2000}
Cox, A. N., 2000, Allen's Astrophysical Quantities, 4th ed., Springer-Verlag, New York
\bibitem[\protect\citeauthoryear{De Marco et al.}{2002}]{deMarco 2002}
De Marco, O., Clayton, G. C., Herwig, F., Pollacco, D. L., Clark, J. S., \& Kilkenny, D., 2002, ApJ, 123, 3387
\bibitem[\protect\citeauthoryear{Grindlay et al.}{2012}]{Grindlay et al. 2012}
Grindlay, J., Tang, S., Los, E., \& Servillat, M. 2012, in New Horizons in Time-Domain Astronomy (IAU Symposium 285), p. 29-34, arXiv:1211.1051
\bibitem[\protect\citeauthoryear{Heck et al.}{1985}]{Heck et al. 1985}
Heck, A., Houziaux, L., Manfroid, J., Jones, D. H. P., \& Andrews, P. J.. 1985, A\&ASupp, 61, 375
\bibitem[\protect\citeauthoryear{Henden \& Munari}{2014}]{Henden and Munari 2014}
Henden, A. \& Munari, U., 2014, Contrib. Astron. Obs. Skalnate Pleso, 43, 518
\bibitem[\protect\citeauthoryear{Herbig}{1964}]{Herbig 1964}
Herbig, G., 1964, ApJ, 140, 1317
\bibitem[\protect\citeauthoryear{Hoffleit}{1930}]{Hoffleit 1930}
Hoffleit, D., 1930, Harvard Bulletin, 874, 1
\bibitem[\protect\citeauthoryear{Hoffleit}{1958}]{Hoffleit 1958}
Hoffleit, D., 1958, AJ, 63, 78
\bibitem[\protect\citeauthoryear{Hoffleit}{1959}]{Hoffleit 1959}
Hoffleit, D., 1959, AJ, 64, 241
\bibitem[\protect\citeauthoryear{Goldsmith et al.}{1990}]{Goldsmith et al. 1990}
Goldsmith, M. J., Evans, A., Albinson, J. S., \& Bode, M. F., 1990, MNRAS, 245, 119
\bibitem[\protect\citeauthoryear{Jeffery et al.}{2001}]{Jeffery et al. 2001}
Jeffery, C. S., Starling, R. L. C., Hill, P. W., \& Pollacco, D., 2001, MNRAS, 321, 111
\bibitem[\protect\citeauthoryear{Jones et al.}{1989}]{Jones et al. 1989}
Jones, K., van Wyk, F., Jeffery, C. S., Marang, F., Shenton, M., Hill, P., \& Westerhuys, J., 1989, South African Astronomical Observatory Circular, 13, 39
\bibitem[\protect\citeauthoryear{Kilkenny et al.}{1985}]{Kilkenny et al. 1985}
Kilkenny, D., Coulson, I. M., Laing, J. D., Jones, J. S., \& Engelbrecht, C., 1985, South African Astronomical Observatory Circular, 9, 87
\bibitem[\protect\citeauthoryear{Landolt}{2009}]{Landolt 2009}
Landolt, A.~U., 2009, AJ, 137, 4186
\bibitem[\protect\citeauthoryear{Marino and Walker}{1971}]{Marina and Walker 1971}
Marino, B. F. \& Walker, W. S. G., 1971, Circular Royal Astron. Soc. New Zealand Var. Star Section, 184, 1
\bibitem[\protect\citeauthoryear{Munari et al.}{2014}]{Munari et al. 2014}
Munari, U., Henden, A., Frigo, A. et al., 2014, AJ, 148, 81
\bibitem[\protect\citeauthoryear{Pandey et al.}{2014}]{Pandey et al. 2014}
Pandey, G., Rao, N. K., Jeffery, C., \& Lambert, D. L., 2014, ApJ, 793, 76
\bibitem[\protect\citeauthoryear{Patat et al.}{1997}]{Patat et al. 1997}
Patat, F., Barbon, R., Cappellaro, E., \& Turatto, M., 1997, A\&A, 317, 423
\bibitem[\protect\citeauthoryear{Pollacco and Hill}{1991}]{Pollacco and Hill 1991}
Pollacco, D. L. \& Hill, P. W., 1991, MNRAS, 248, 572
\bibitem[\protect\citeauthoryear{Rao et al.}{1993}]{Rao et al. 1993}
Rao, N. K., Giridhar, S., \& Lambert, D. L., 1993, A\&A, 290, 201
\bibitem[\protect\citeauthoryear{Rao et al.}{2012}]{Rao et al. 2012}
Rao, N.~K., Lambert, D. L., Garcia-Hernandez, D. A., Jeffery, C. S., Woolf, V. M., \& McArthur, B., 2012, ApJLett, 760, L3
\bibitem[\protect\citeauthoryear{Rao et al.}{2013}]{Rao et al. 2013}
Rao, N.~K., Lambert, D. L., Garcia-Hernandez, D. A., \& Manchado, A., 2013, MNRAS, 431, 159
\bibitem[\protect\citeauthoryear{Saio}{1988}]{Saio 1988}
Saio, H., 1988, MNRAS, 235, 203
\bibitem[\protect\citeauthoryear{Sandage}{2001}]{Sandage 2001}
Sandage, A., 2001, PASP, 113, 267
\bibitem[\protect\citeauthoryear{Schaefer}{1981}]{Schaefer 1981}
Schaefer, B.~E., 1981, PASP, 93, 253
\bibitem[\protect\citeauthoryear{Schaefer}{1994}]{Schaefer 1994}
Schaefer, B.~E., 1994, ApJ, 426, 493
\bibitem[\protect\citeauthoryear{Schaefer}{1995}]{Schaefer 1995}
Schaefer, B.~E., 1995, ApJLett, 447, L13
\bibitem[\protect\citeauthoryear{Schaefer}{1996}]{Schaefer 1996}
Schaefer, B.~E., 1996, ApJLett, 460, L19
\bibitem[\protect\citeauthoryear{Schaefer}{1998}]{Schaefer 1998}
Schaefer, B.~E., 1998, ApJ, 509, 80
\bibitem[\protect\citeauthoryear{Schaefer and Edwards}{2015}]{Schaefer and Edwards 2015}
Schaefer, B.~E. \& Edwards, Z. I., 2015, ApJ, 812, 133
\bibitem[\protect\citeauthoryear{Sherwood}{1975}]{Sherwood 1975}
Sherwood, V. E., 1975, in Variable Stars and Stellar Evolution, ed. V. E. Sherwood and L. Plaut (Dordrecht: Reidel), 147
\bibitem[\protect\citeauthoryear{Soszynski et al.}{2009}]{Soszynski et al. 2009}
Soszynski, I., Udalski, A., Szymanski, M. K. et al., 2009, Acta Astron., 59, 335
\bibitem[\protect\citeauthoryear{Woods}{1928}]{Woods 1928}
Woods, I. E., 1928, Bulletin Harvard College Obs., 855, 22
\end{thebibliography}



\bsp	
\label{lastpage}
\end{document}